\documentclass[conference]{IEEEtran}

\usepackage{graphicx}
\graphicspath{{figs/}{../figs/}}
\usepackage{amsmath}
\usepackage{amssymb}
\usepackage{array}
\usepackage{hyperref}
\usepackage{booktabs}
\usepackage{multirow}
\usepackage{iftex}
\usepackage{pgfplots}
\usepgfplotslibrary{groupplots}
\pgfplotsset{compat=1.18}

 \usepackage{xspace}
  \newcommand{\chordtype}[1]{\texttt{#1}\xspace}
  \newcommand{\chord}[1]{\texttt{#1}\xspace}
  \newcommand{\rn}[1]{\textup{#1}\xspace}
   \usepackage{xspace}
  \newcommand{\key}[2]{\textup{#1~#2}\xspace} 

\ifPDFTeX
  \usepackage{newtxtext,newtxmath}
\else
  \usepackage{fontspec}
  \defaultfontfeatures{Ligatures=TeX}
\fi

\title{Structure-Aware Piano Accompaniment via Style Planning and Dataset-Aligned Pattern Retrieval}

\author{
	    Wanyu Zang \\
	    Computer Science \\
	    Lewis University \\
	    \texttt{wzang@lewis.edu}
	    \and
	    Yang Yu \\
	    Nexus Art Foundation \\
	    \texttt{yuy.cse@gmail.com}
	    \and
	    Meng Yu \\
	    Computer Science \\
	    Governors State University \\
	    \texttt{myu.mengyu@gmail.com}
}

\begin{document}

\maketitle

\begin{abstract}
We introduce a structure-aware approach for symbolic piano accompaniment that decouples high-level planning from note-level realization. A lightweight transformer predicts an interpretable, per-measure style plan conditioned on section/phrase structure and functional harmony, and a retriever then selects and reharmonizes human-performed piano patterns from a corpus. We formulate retrieval as pattern matching under an explicit energy with terms for harmonic feasibility, structural-role compatibility, voice-leading continuity, style preferences, and repetition control. Given a structured lead sheet and optional keyword prompts, the system generates piano-accompaniment MIDI. In our experiments, transformer style-planner-guided retrieval produces diverse long-form accompaniments with strong style realization. We further analyze planner ablations and quantify inter-style isolation. Experimental results demonstrate the effectiveness of our inference-time approach for piano accompaniment generation.
\end{abstract}

\begin{IEEEkeywords}
Piano accompaniment, retrieval-based generation, style control, functional harmony, structure-aware modeling, transformers
\end{IEEEkeywords}

\section{Introduction}
\label{sec:introduction}
Automatic piano accompaniment generation requires coordinated control of harmony, rhythm, texture, and long-range form. Recent neural approaches generate notes directly with Transformers or diffusion models \cite{huang2019music,oore2020time,wang2024wholesong,min2023polyffusion}, but long-form controllability remains difficult. In particular, small symbolic datasets exacerbate mode collapse, and models often entangle global form with local realization. Retrieval-based systems can reuse human-written patterns, but they still need reliable high-level planning and diversity-aware selection to avoid repetitive outputs \cite{zhao2021accomontage,yi2022accomontage2}.

In this paper, we decouple style learning and planning from note-level pattern and texture realization. Instead of synthesizing every note, we \emph{plan} accompaniment style and \emph{retrieve} compatible patterns from POP909 \cite{pop909}, which was performed by real pianists. Pattern retrieval is attractive because every pattern comes from a musician. We preserve articulation cues as much as possible, including onset and offset timing, duration offsets, and velocity. This produces realistic results.

This setting raises several challenges. POP909 \cite{pop909} contains only 909 songs, and we further split the dataset into training and validation sets. As a result, the Transformer sees limited data. We must therefore choose carefully what the Transformer should learn and output. In addition, patterns in the dataset serve different roles and functions, and they represent different styles. For example, a pattern or texture for a pickup beat or an ending is usually not a good candidate for a measure in the middle of a phrase. Note durations can also cross bar lines due to anticipation. Therefore, accompaniment generation is not a simple puzzle of stitching patterns together under direct Transformer guidance. Because the dataset is small, it is not always possible to find candidates that satisfy role constraints, harmonic constraints, and style preferences at the same time.

Music theory, keyboard voicing, and other constraints further restrict what we can select from the dataset. For example, \chord{C:maj7} can serve as \rn{I} in \key{C}{major} but as \rn{V} in \key{F}{major}. Different functions imply different progressions and voice-leading choices in the dataset. Therefore, matching is not a simple string search for \chord{C:maj7}.

In this paper, we propose a Transformer-guided style planner and retriever framework. We also describe a retriever strategy that follows the Transformer guidance while producing dataset-grounded realizations that preserve the rhythmic, registral, and voicing characteristics of real pianist performances.

Our main contributions are as follows.
\begin{itemize}
   \item \textbf{Dataset key and mode normalization to effectively use patterns based on harmony function:} We normalize all songs to \key{C}{major} or \key{A}{minor} and transpose chords accordingly. We annotate chords with Roman numerals and use chord functions for analysis and retrieval. This improves dataset utilization and pattern matching.
    \item \textbf{Style planning as a learned intermediate representation:} We train a compact Transformer encoder to impute masked, measure-level style slots from structure and local context. This supports controllable section contrast.
    \item \textbf{Functional, structure-aware retrieval for accompaniment:} We build a dataset index and a matching procedure that combines style signatures with factorized harmony (including transpose-invariant Roman-numeral features) and supports reharmonization.
    \item \textbf{Long-form diversity analysis and tuning:} Our experiments show that our final pattern diversity aligns with dataset statistics while maintaining strong section-level pattern contrast.
\end{itemize}

This paper is organized as follows. Section~\ref{sec:introduction} motivates the problem and summarizes our contributions. Section~\ref{sec:relatedwork} reviews related work. Section~\ref{sec:design} provides an overview of our system design. Section~\ref{sec:planner} details the Transformer style planner. Section~\ref{sec:retriever} describes dataset pattern retrieval and adaptation. Section~\ref{sec:experiments} reports experiments and results. Section~\ref{sec:discussion} discusses findings and limitations. Section~\ref{sec:futurework} outlines future directions, and Section~\ref{sec:conclusion} concludes.

\section{Related Work}
\label{sec:relatedwork}

Symbolic accompaniment and arrangement systems differ along two axes. The first is how note-level realization is produced (learned note/event generators vs.\ retrieval and templates). The second is how long-form style and structure control is exposed. We discuss representative approaches below.
Autoregressive models, including \cite{huang2019music} and \cite{oore2020time}, and diffusion models, including \cite{min2023polyffusion} and \cite{wang2024wholesong}, synthesize note or event sequences directly. Conditional accompaniment and infilling can be framed as conditioning or inpainting \cite{thickstun2024anticipatory}, and whole-song systems can additionally model explicit long-term structure via hierarchical representations \cite{wang2024wholesong}. These approaches are typically trained on large-scale datasets such as Lakh \cite{raffel2016learning}. They expose control mainly through conditioning signals or learned latent codes, rather than an intermediate, musician-interpretable style plan aligned with section and phrase roles. Related work also improves event representations for stronger metrical conditioning, e.g., the REMI representation used by Pop Music Transformer \cite{huang2020popmusictransformer}. In our work, we use a much smaller dataset (909 songs), so we instead adopt a retrieval-first approach.

Recent arrangement and re-orchestration models factorize style and content to improve multi-track coherence and controllability. For example, Zhao et al.~\cite{zhao2024structured} propose a two-stage system that retrieves piano texture styles from a lead sheet; the second stage generates orchestration with a learned style prior. METEOR~\cite{le2025meteor} uses a transformer VAE for melody-aware, texture-controllable symbolic re-orchestration. Unlike our approach, these works focus on orchestration and involve multiple instruments, whereas we focus on piano accompaniment only.

Retrieval-based systems reuse performed patterns to obtain realistic voicings, articulations, rhythms, and textures without training a note generator. Accomontage \cite{zhao2021accomontage} assembles phrase-level templates with dynamic programming and then reharmonizes them via style transfer. AccoMontage2 \cite{yi2022accomontage2} extends this line by adding a full-song harmonization module (generating chord progressions from a lead melody) and offering user-facing controls over harmonic and texture styles.

Our work differs in both interface and granularity. We assume a lead sheet with functional harmony and explicit section/phrase cues as input, and focus on learning a per-measure, musician-interpretable style plan that guides measure-level retrieval under an explicit energy model. This enables structure-aware local variation and continuity control without relying on phrase-level dynamic programming or a separate harmonization stage.

Beyond lead-sheet-conditioned arrangement, piano cover generation systems such as PiCoGen \cite{tan2024picogen} factor the task into transcribing a lead sheet from audio and then generating a symbolic piano cover. In contrast, we study structure-aware accompaniment generation given a symbolic lead sheet (optionally with inferred structure cues), and emphasize dataset-aligned pattern retrieval for realistic pianistic textures.

Relative and functional representations such as Roman numerals improve generalization across keys and mitigate data scarcity \cite{huang2024emotion}, aligning with long-standing theory that emphasizes functional roles over absolute pitch \cite{schoenberg1983theory}. In our setting, key-invariant functional harmony supports planning and retrieval across transpositions.

Auto-accompaniment is widely deployed in commercial tools. Pattern-library systems such as Band-in-a-Box \cite{bandinabox} and Toontrack EZkeys \cite{ezkeys2} select and adapt curated MIDI patterns from chord progressions, and DAWs provide related workflows (e.g., Logic Pro \cite{logicpro} and Pro Tools \cite{protools}); however, these systems are proprietary and hard to evaluate reproducibly. End-to-end audio generation systems such as Suno \cite{suno} and Udio \cite{udio} can produce complete songs, but they typically do not provide transparent, structure-conditioned symbolic control.

\begin{table*}[t]
\centering
\setlength{\tabcolsep}{1pt}
\caption{Representative comparisons along key axes for symbolic accompaniment and arrangement. ``Structure'' indicates whether explicit section and phrase cues are used rather than implicit or learned structure. ``Planning'' indicates whether a distinct song-level plan is produced before note-level realization.}
\label{tab:system_comparison}
{\small
\begin{tabular}{@{} >{\raggedright\arraybackslash}p{0.28\linewidth} >{\raggedright\arraybackslash}p{0.16\linewidth} c >{\raggedright\arraybackslash}p{0.16\linewidth} >{\raggedright\arraybackslash}p{0.14\linewidth} >{\raggedright\arraybackslash}p{0.12\linewidth}@{}}
\toprule
Approach & Realization & Retr. & Structure & Style & Planning \\
\midrule
Autoregressive note-level \cite{huang2019music} and \cite{oore2020time} & Note or event gen. & N & Implicit & Priming/cond. & None \\
Beat-based modeling (REMI) \cite{huang2020popmusictransformer} & Note/event gen. & N & Implicit & Representation & None \\
Anticipatory infilling \cite{thickstun2024anticipatory} & Note or event gen. & N & Control events & Control tokens & None \\
Diffusion inpainting/controls \cite{min2023polyffusion} & Note or event gen. & N & Local/inpaint & Conditioning & None \\
Whole-song hierarchical diffusion \cite{wang2024wholesong} & Note or event gen. & N & Explicit hierarchy & Conditioning & Hierarchical \\
Style-prior arrangement \cite{zhao2024structured} & Hybrid retr.+gen. & Y & Phrase and section & Style priors & Transformer prior \\
Texture-controllable re-orchestration \cite{le2025meteor} & VAE transfer & N & Bar and track & Texture ctrl & None \\
Retrieval + DP templates \cite{zhao2021accomontage} & Phrase templates & Y & Phrase-level & Style transfer & DP \\
AccoMontage2 \cite{yi2022accomontage2} & Harmonize+arrange & Y & Phrase-level & User controls & DP+pipeline \\
PiCoGen \cite{tan2024picogen} & Transcribe+cover gen. & N & Implicit & Conditioning & Two-stage \\
Ours & Measure retrieval & Y & Section and phrase & Keywords to slots & Transformer \\
\bottomrule
\end{tabular}
}
\end{table*}

Table~\ref{tab:system_comparison} summarizes how our design differs from representative prior approaches. In particular, compared to Accomontage \cite{zhao2021accomontage}, our method introduces an explicit, musician-interpretable style-planning layer and performs measure-level retrieval under structured constraints, enabling fine-grained section/phrase control in the small-data regime.

In summary, our design combines an explicit, transformer-based, structure-aware style planner that produces a musician-interpretable intermediate plan with measure-level retrieval under explicit constraints. This enables controllable long-form accompaniment in a POP909-scale setting without training a note-level generator.

\paragraph{Why we do not benchmark against note-level generators.}
Direct note-level accompaniment generators (e.g., autoregressive Transformers or diffusion models) are typically trained on substantially larger symbolic corpora and evaluated with frame-level or onset-based metrics. In contrast, our work targets a different operating regime: controllable, long-form piano accompaniment from a small, curated dataset (POP909) with explicit structural roles.

In preliminary experiments, we observed that note-level models trained on POP909 can exhibit severe repetition and structural drift under free generation, despite strong teacher-forced metrics---an issue widely reported in prior work. As our goal is not to improve note-level likelihood but to enable interpretable, structure-aware planning with dataset-grounded realizations, we focus evaluation on diversity, retrieval feasibility, and long-form behavior rather than direct note-wise accuracy.

We therefore position our system as complementary to note-level generators rather than a drop-in replacement, and leave large-scale note-level benchmarking to future hybrid extensions.

\section{Design Overview}
\label{sec:design}

\subsection{Architecture}
\label{subsec:architecture}

\begin{figure*}[t]
\centering
\includegraphics[width=0.75\textwidth]{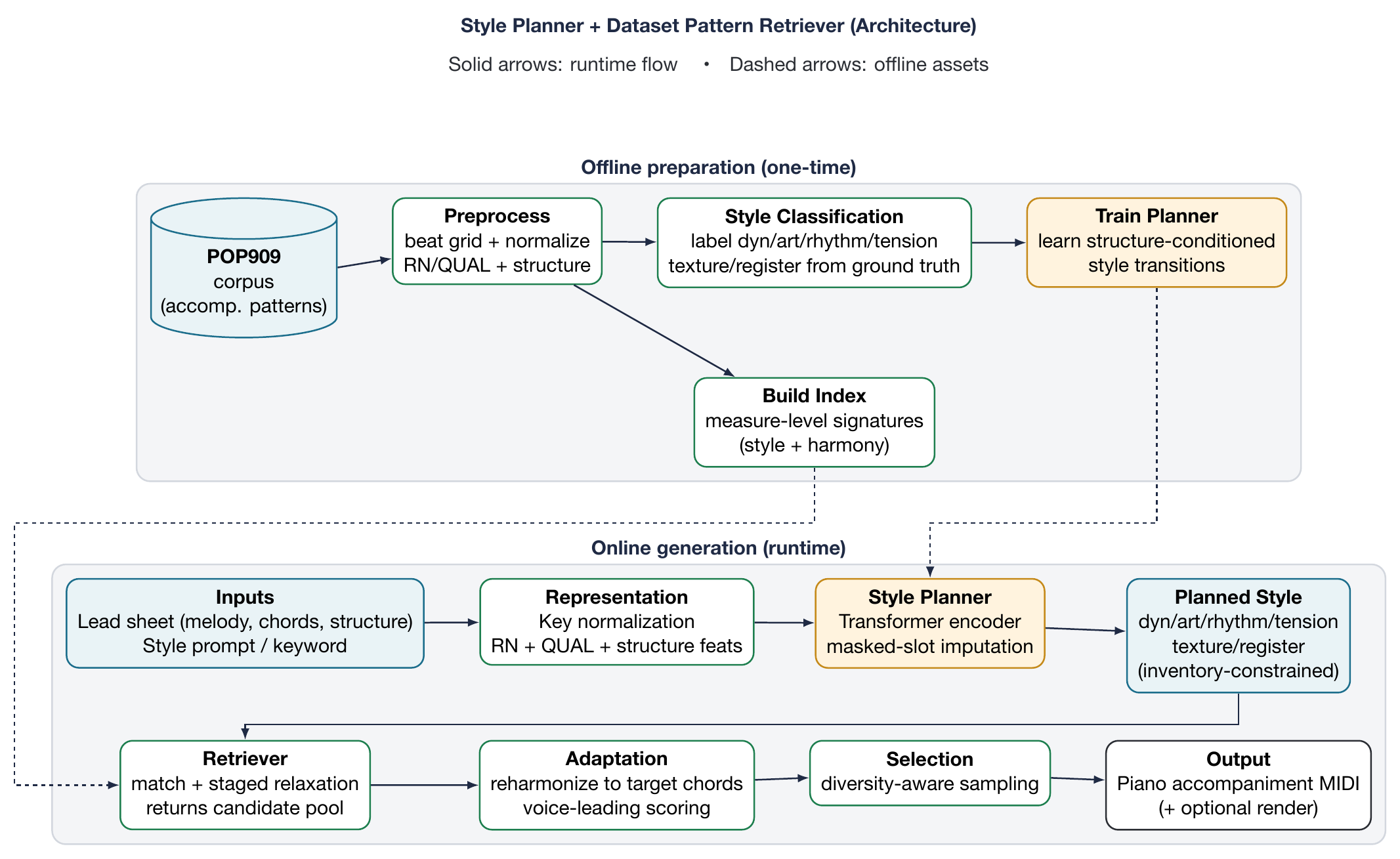}
\caption{Overview of our style-planning and retrieval-based accompaniment pipeline. Offline preparation derives discrete measure-level style labels (dynamics, articulation, rhythm, tension, texture, register) and builds a measure-level index from POP909. Online generation predicts per-measure style slots with a Transformer style planner, then retrieves and adapts dataset patterns to the target harmony with voice-leading-aware selection. Solid arrows indicate runtime flow; dashed arrows indicate offline assets.}
\label{fig:architecture}
\end{figure*}

Fig.~\ref{fig:architecture} provides an overview of the proposed pipeline. We map each song to a unified, beat-synchronous representation and factorize harmony so that all modules interoperate through a shared set of data structures. The Transformer style planner, pattern retriever, renderer, and supporting components consume and produce this common representation.
Our design philosophy is to decouple high-level, structure-conditioned \emph{style planning} from note-level \emph{pattern realization}.

Our system operates in a key-normalized functional space (Roman numerals and factorized chord quality) and uses explicit structure labels (section and phrase roles) to provide long-range form cues. POP909 includes song-structure annotations for all dataset pieces; to support songs outside the dataset, we infer analogous structure cues with All-In-One~\cite{all-in-one-paper} and combine them with the lead sheet for planning and retrieval. The Transformer style planner predicts per-measure \emph{style slots}---dynamics, articulation, rhythm, tension (syncopation), texture, and register---from structure and local context, optionally biased by user keywords. A dataset pattern retriever then selects measures matching the planned style and chord-quality context using voice-leading- and diversity-aware scoring, and reharmonizes them to the target harmony while preserving rhythm and voicing.

\subsection{Dataset and Representation}

POP909 \cite{pop909} is a carefully curated dataset of piano performances of Chinese pop songs with rich annotations. Each MIDI file includes aligned beat locations, chord labels and change points, phrase boundaries, and song-level section structure, among other metadata.

From the raw POP909 MIDI files and annotations, we build a single structured representation that supports both learning and retrieval. The same preprocessed artifact serves (i) as the training dataset for our Transformer style planner and (ii) as the retrieval corpus from which we index and retrieve real piano accompaniment patterns. Throughout the paper, we use ``dataset'' when discussing training/evaluation and ``corpus'' when describing retrieval, although both refer to this same POP909-derived resource.

In our work, we use POP909 \cite{pop909} to extract per-measure accompaniment styles and patterns. Each song is normalized to \key{C}{major} or \key{A}{minor} to preserve mode, while detecting and preserving modulations. The normalization emphasizes chord \emph{function} rather than chord \emph{name}: the same chord can serve different functions under different keys, leading to different roles in a progression and, consequently, different voicings and textural continuations across measures. During preprocessing, we extract key along with factorized chord quality (QUAL) and optional extensions/alterations.

This key-invariant representation captures these functional distinctions without tying them to absolute chord symbols. It supports transpose-aware matching and reduces sparsity when conditioning patterns on harmony.

We encode global structure with section labels (e.g., intro, verse, chorus, etc.) and phrase roles derived from POP909's structure annotations. To make retrieval controllable, we summarize accompaniment intent at the measure level with discrete \emph{style slots}: dynamics, articulation, rhythm, tension, texture, and register. We derive these labels from the ground-truth accompaniment using a feature-based style labeler (velocity, articulation, onset patterns, syncopation, and register cues) and construct an \emph{inventory} of slot combinations observed in the training dataset. During planning and inference, we constrain predictions to this inventory to avoid requesting styles that do not exist in the retrieval database.

Our style analyzer classifies accompaniment styles per measure based on dynamics, articulation, rhythmic activity, syncopation, texture scores, and register statistics. These labels supervise the Transformer planner and serve as retrieval keys during generation.

\section{Transformer Style Planner}
\label{sec:planner}

We cast accompaniment generation as a two-stage decision process: (i) plan an interpretable, measure-level ``style trajectory'' and (ii) retrieve and adapt concrete piano patterns from the dataset to realize that plan. The Transformer style planner is responsible only for stage (i). Rather than emitting notes, it predicts discrete per-measure style slots that closely match how pianists reason about accompaniment in practice: dynamics (how forceful), articulation (how percussive vs.\ legato), rhythmic activity, syncopation/tension, texture family (block/arpeggio/Alberti/stride/etc.), and register focus. This division of labor makes the planner data-efficient and musically transparent, while allowing the downstream retriever to enforce idiomatic pianistic realization.

\paragraph{Problem Formulation.}
For each measure $t$, we define a style slot vector
\begin{equation}
s_t = \left(y^{\text{dyn}}_t, y^{\text{art}}_t, y^{\text{rhythm}}_t, y^{\text{tension}}_t, y^{\text{texture}}_t, y^{\text{register}}_t\right),
\end{equation}
where each component is a categorical label predicted at the measure level. The planner models a conditional distribution $p_{\theta}(s_t \mid X_{t-k:t+k}, z_t, u_t)$, where $X_{t-k:t+k}$ is a local context window of conditioning tokens centered at measure $t$ (including global song tokens and nearby measures), $z_t$ is a compact structure vector encoding section/phrase context and within-section position, and $u_t$ is an optional style-prompt vector derived from user keywords (or \texttt{AUTO} when no preference is given). In the final system we primarily train and use a \emph{style-only} variant that predicts these six measure-level style slots, since they form the explicit interface to retrieval and capture musically salient accompaniment variation with a small, data-efficient output space.

\paragraph{Input Representation (Conditioning Tokens + Structure Vector).}
The conditioning sequence is a symbolic tape that summarizes the lead sheet and its functional harmonic context without expanding into note-level event streams. Concretely, tokens include global context (key, tempo bin, meter), measure-level structure labels (section and phrase markers), and harmonic descriptors expressed in a transposition-invariant functional space (Roman numeral and chord quality, with additional descriptors such as bass degree, extensions, and inversion when available). We also include lightweight melodic summaries (e.g., density, register, rhythm template, contour) that capture how much surface activity the melody demands without forcing the planner to model exact note sequences. The planner is trained on windows spanning $k=8$ past measures and one future measure, which is musically motivated: accompanists strongly condition on what just happened (texture inertia, continuation of rhythmic motor) while also anticipating near-future harmonic events (cadences, section boundaries). Since the lead sheet is known at inference time, using a bidirectional context window is appropriate and avoids the instability of autoregressive note generation.

In addition to tokens, each target measure receives a continuous structure vector $z_t$ that encodes (i) a one-hot section label, (ii) a one-hot phrase-role label (begin/mid/end/single), (iii) three normalized positional scalars (beat index within measure, measure index within section, measure index within phrase), and (iv) a one-hot dynamics-trend label (none/crescendo/decrescendo/swell/drop). This vector injects long-range organizational information (section function and phrase position) that is difficult to infer reliably from local harmony alone, and it mirrors how pianists shape accompaniment: texture often evolves systematically across phrases (opening sparse, intensifying toward cadences) and differs by section role (verse vs. chorus). When style prompting is enabled, we concatenate a soft keyword-derived prompt vector $u_t$ to $z_t$ so the model can condition on both structural intent and explicit stylistic preferences.

\paragraph{Transformer Design and Rationale.}
We implement the planner as a compact Transformer encoder \cite{vaswani2017attention} trained with a masked-slot objective (BERT-style imputation \cite{devlin2019bert}). The default configuration uses $d_{\text{model}}=256$, $L=4$ encoder layers, $H=8$ attention heads, feedforward width $d_{\text{ff}}=1024$. This scale is deliberate: POP909 provides strong stylistic diversity but only 909 songs, and our planner's output space remains small (six categorical slots per measure). A compact bidirectional encoder therefore matches both the supervision regime and the role of the planner in the pipeline: it must be reliable, data-efficient, and interpretable, not expressive enough to memorize note-level realizations. The planner's attention operates over the entire token window, enabling it to fuse harmonic function, local repetition, and structural markers into a coherent per-measure style decision.

\paragraph{Measure-Level Pooling and Multi-Head Slot Prediction.}
Because the input is a long token sequence spanning multiple measures, we explicitly anchor predictions to the target measure. After encoding the token window into contextual embeddings, we pool (mean-average) embeddings over the token span corresponding to the current measure to obtain a measure representation $h_t$. We then fuse structure and prompt information via an additive projection. Let $\bar{z}_t = [z_t;u_t]$ denote the concatenation of structure features and the (optional) style prompt; we compute $\tilde{h}_t = h_t + W_z \bar{z}_t$, and predict each slot with its own classification head (multi-task learning): 
\begin{equation}
p_{\theta}(y^j_t \mid \cdot) = \text{softmax}(W_j \tilde{h}_t), \quad j\in\mathcal{S}.
\end{equation}
Separate heads allow the model to represent musically meaningful asymmetries: for instance, texture and register choices are often section- and harmony-sensitive, while rhythm and tension relate more directly to surface activity and syncopation, and dynamics/articulation reflect performance intent.

\paragraph{Training Objective (Masked Style Slots).}
Training mirrors inference: for each training sample we mask the accompaniment-derived style tokens in the current measure only (replacing them with padding and masking them from attention), while keeping the surrounding measures intact. The model is optimized with cross-entropy losses on the masked slots only. This forces the planner to (a) infer style from musical context when no explicit preference is provided and (b) learn realistic transitions by observing unmasked neighboring measures. In practice, this objective encourages stable style inertia across adjacent measures (common in real accompaniment), while still allowing sharp changes at structurally meaningful boundaries (phrase starts, chorus entries) when supported by $z_t$ and harmonic/melodic cues.

\paragraph{Loss Function.}
Let $\mathcal{S}$ denote the set of predicted style slots (dyn, art, rhythm, tension, texture, register), and let $y_{t,j}$ be the ground-truth class for slot $j \in \mathcal{S}$ at measure $t$. For each slot head we obtain logits $\ell_{t,j}$ and a categorical distribution $\hat{p}_{t,j}=\mathrm{softmax}(\ell_{t,j})$. We optimize a masked multi-task cross-entropy over the set of masked, labeled slots $\mathcal{M}$:
\begin{equation}
\mathcal{L}_{\text{plan}}(\theta) = \frac{1}{|\mathcal{M}|}\sum_{(t,j)\in\mathcal{M}}\mathrm{CE}(\hat{p}_{t,j}, y_{t,j}).
\end{equation}
Unmasked slots are excluded from $\mathcal{M}$ to prevent trivial copying, and slots without valid labels are ignored. In the style-only setting, $\mathcal{M}$ includes the six style slots for the current measure, so $\mathcal{L}_{\text{plan}}$ becomes the mean per-slot cross-entropy; when training a richer planner that also predicts beat-level or non-style slots, we extend the same masked-loss definition to those heads.

\paragraph{Supporting Style Preferences and Keyword Prompts.}
We support user-controllable style preferences without requiring free-form text modeling. Users specify a small set of musician-friendly keywords (15 in our current implementation; e.g., \texttt{quiet}, \texttt{gentle}, \texttt{busy}, \texttt{sparse}, \texttt{fast}, \texttt{alberti}, \texttt{stride}). Each keyword is mapped to slot-wise \emph{allowed}, \emph{preferred}, and \emph{excluded} label sets over the six style axes (dyn/art/rhythm/tension/texture/register), reflecting pianistic semantics: some keywords specify a texture family directly, while others constrain performance intent (dynamics, articulation), rhythmic activity, and syncopation. Multiple keywords are merged by intersecting allowed sets (when specified), unioning preferred labels, and unioning exclusions.

\emph{Soft prompt conditioning.} From a keyword set $\kappa$, we build a continuous prompt vector $u(\kappa)\in\mathbb{R}^{29}$ by concatenating per-axis distributions over discrete labels (including an \texttt{AUTO} label on each axis). For each axis $a$ with label set $\mathcal{Y}_a$, we define nonnegative weights
\begin{equation}
w_a(y;\kappa)=
\begin{cases}
1.0 & y\in \mathcal{P}_a(\kappa) \\
0.5 & y\in \mathcal{A}_a(\kappa)\setminus \mathcal{P}_a(\kappa) \\
0.0 & y\in \mathcal{E}_a(\kappa) \\
0.0 & \text{otherwise},
\end{cases}
\end{equation}
where $\mathcal{A}_a,\mathcal{P}_a,\mathcal{E}_a$ are the merged allowed/preferred/excluded sets induced by $\kappa$. If $\sum_{y\in\mathcal{Y}_a} w_a(y;\kappa)=0$ (no active constraint survives), we set $w_a(\texttt{AUTO};\kappa)=1.0$. We then normalize within each axis, $u_a(y\mid\kappa)=w_a(y;\kappa)/\sum_{y'} w_a(y';\kappa)$, and concatenate $u(\kappa)=[u_a]_{a}$. Intuitively, $u(\kappa)$ provides a \emph{soft prior over style labels} rather than a hard instruction, letting the planner trade off preference satisfaction against contextual plausibility.

\emph{During training}, when prompt conditioning is enabled we append $u_t=u(\kappa_t)$ to the structure input for the current measure. To avoid inconsistent conditioning while still teaching controllability, we sample $\kappa_t$ from keywords that are \emph{compatible} with the ground-truth style labels (i.e., do not exclude the target and respect any required sets) and mix prompt scopes (song-level and section-level keywords for stability, and occasional measure-level keywords for variation). With a configurable probability, we drop the entire prompt to \texttt{AUTO}-only, forcing the model to remain capable of inferring reasonable defaults from harmony and structure when prompts are absent or underspecified.

\emph{During inference}, user keywords (global or section-specific) produce the same prompt vector $u_t$ and bias the planner through the augmented structure input. For reliability and retrieval feasibility, we also carry the induced constraints into the inventory projection stage: candidate joint style vectors that violate allowed/excluded sets are filtered or penalized (with staged relaxation when prompts over-specify the corpus), and section anchors can be used to encourage within-section coherence. This combination makes keyword control musically meaningful while guaranteeing that the final planned style remains realizable by the dataset retriever.

\paragraph{Inventory-Constrained Decoding as Projection.}
The planner predicts each slot marginally, but the downstream retriever requires a joint style vector that is both musically plausible and present in the dataset. We therefore ``project'' model predictions onto the empirical inventory $\mathcal{I}_{\ell}$ of valid slot vectors observed for measure length $\ell$ by selecting a maximum a posteriori (MAP) candidate:
\begin{equation}
\hat{s}_t = \arg\max_{s\in\mathcal{I}_{\ell_t}}
\Bigg[
\sum_{j\in\mathcal{S}} \log \hat{p}_{t,j}(s_j)
+ \lambda_{\pi}\log \pi_{\ell_t}(s)
\Bigg]
\end{equation}
where $\pi_{\ell}(s)$ is the empirical frequency of style vector $s$ in the training corpus (and $\lambda_{\pi}$ controls the strength of this prior). When $\lambda_{\pi}=0$, the selection reduces to maximum likelihood under the planner; with $\lambda_{\pi}>0$, it prefers corpus-typical realizations when the model is uncertain. This formulation clarifies why a log-count prior is effective: it implements a simple, dataset-grounded regularizer without training any note-level model.

\paragraph{Section-Level Coherence and Contrast.}
To encourage long-form coherence, we can introduce a section anchor $a_{\text{sec}}$ (a representative style vector for each section label) and penalize deviations within the section while allowing controlled contrast between sections. One convenient form is a Hamming-distance penalty over the slot dimensions:
\begin{equation}
\begin{split}
\hat{s}_t = \arg\max_{s\in\mathcal{I}_{\ell_t}} \Bigg[
\sum_{j\in\mathcal{S}} \log \hat{p}_{t,j}(s_j)
+ \lambda_{\pi}\log \pi_{\ell_t}(s) \\
- \lambda_{a} \, d_H\!\left(s, a_{\text{sec}(t)}\right)
\Bigg]
\end{split}
\end{equation}
where $d_H$ counts how many slot components differ. Musically, the anchor acts like a pianist's ``accompaniment concept'' for the section (e.g., verse = sparse Alberti, chorus = denser block/arp), while the planner still chooses measure-level variations around that concept.

\paragraph{Keyword Control as Constrained Optimization.}
In our system, keywords induce both a \emph{conditioning signal} (the soft prompt vector $u_t$) and a \emph{feasible-region view} over styles. Given a keyword set $\kappa$, we define a constraint set $\mathcal{C}(\kappa)\subseteq\mathcal{I}_{\ell}$ as the subset of inventory style vectors whose slot labels satisfy the merged allowed/excluded rules. Hard prompting corresponds to restricting the inventory projection to $\mathcal{I}_{\ell_t}\cap\mathcal{C}(\kappa)$; soft prompting corresponds to adding a penalty for violations (or rewarding preferred labels) within the joint selection objective, which is especially useful when $\mathcal{C}(\kappa)$ is small or empty. When constraints over-specify the dataset, we use staged relaxation to guarantee feasibility while preserving as much of the prompt intent as possible (e.g., relaxing texture before dynamics).

\paragraph{Context Window, Anticipation, and Style Dynamics.}
Although the planner is not autoregressive in notes, it models a structured style process. The bidirectional Transformer encoder over a local window can be viewed as learning a high-order, context-conditioned transition model: recent measures provide inertia and continuity (the accompanist keeps the same motor and register), while a short lookahead enables anticipatory adjustments near cadences and section boundaries. This aligns with pianistic practice, where accompaniment often ``prepares'' a boundary by changing texture, register focus, or rhythmic activity before the formal downbeat of a new section.

\paragraph{Stochastic vs. Deterministic Planning.}
Because the planner outputs categorical distributions over each slot, inference can be performed either deterministically (MAP/argmax) or stochastically by sampling with a temperature (optionally with top-$k$ or top-$p$ truncation). We support sampling both when selecting a joint vector from the inventory (sampling over candidate scores) and when falling back to per-slot prediction, providing a practical diversity knob without changing the planner.

\paragraph{Interpretability and Retrieval Feasibility.}
The discrete slot interface makes the planner interpretable and evaluable: each decision corresponds to a recognizable accompaniment attribute, and failures can be diagnosed at the level of dynamics/articulation/rhythm/tension/texture/register rather than as opaque note errors. Moreover, constraining outputs to $\mathcal{I}_{\ell}$ enforces retrieval feasibility by construction, preventing the planner from requesting out-of-distribution combinations that the dataset cannot realize. This division of labor is particularly important in the small-data regime: the planner learns only the structure-conditioned style dynamics, while the dataset provides the detailed pianistic realizations.

\paragraph{Why Learned Style Planning is Necessary.}
While heuristic rules can capture local accompaniment conventions, they struggle to model longer-range dependencies such as gradual texture buildup, anticipatory changes before section boundaries, or consistent contrast between verse and chorus across an entire song. The Transformer style planner learns these patterns directly from data by conditioning on both local harmonic context and explicit structural cues, enabling coherent style trajectories that would be difficult to hand-design. Importantly, the planner operates over a compact, interpretable output space, making it data-efficient and compatible with retrieval-based realization.

In summary, the Transformer style planner is intentionally small, bidirectional, and slot-based: it converts lead-sheet functional context and structural cues into a discrete, interpretable plan that can be steered by musician-friendly keywords and safely grounded in the corpus via inventory constraints.

\section{Dataset Pattern Retriever}
\label{sec:retriever}

The style planner predicts a discrete, measure-level intent, but the retriever is responsible for turning that intent into concrete pianistic realization.
Our retriever is \emph{dataset-first}: it selects a measure from a corpus of human-performed piano accompaniment and applies a lightweight harmonic adaptation step, rather than decoding notes from a neural generator.
This design emphasizes (i) idiomatic texture and timing, (ii) explicit constraints that reflect musical practice, and (iii) predictable behavior in the small-data regime.

\paragraph{Beyond Nearest-Neighbor Retrieval.}
Although our system retrieves accompaniment measures from a dataset, retrieval is not performed via static similarity or nearest-neighbor matching. Instead, candidate selection is formulated as a constrained optimization problem that jointly considers harmonic feasibility, structural role compatibility, voice-leading continuity, and long-form diversity control. The retriever further employs staged relaxation and slack budgets to guarantee feasibility under small candidate pools, enabling graceful degradation rather than brittle failure. This design allows retrieval to function as a controllable realization mechanism guided by a learned style plan, rather than as a memory lookup alone.

\paragraph{Problem Formulation.}
Let $\mathcal{D}$ be the set of accompaniment measures in the retrieval corpus.
Each candidate measure $m\in\mathcal{D}$ carries (a) a symbolic performance pattern (multi-voice onsets, sustains, and expressive attributes), (b) a style descriptor $\mathbf{s}(m)$ (our six style slots), (c) a local harmony descriptor $\mathbf{q}(m)$ (a per-beat chord-quality sequence), and (d) a structural role label $\mathbf{r}(m)$ (phrase start/mid/end, boundary cues).
For each target measure $t$, we are given the planned style $\mathbf{s}_t$ from the Transformer planner and the lead-sheet harmony context $\mathbf{q}_t$ (plus global key/mode when available).
The retriever chooses a source measure and produces an adapted accompaniment:
\begin{equation}
m_t^{*} \;=\; \arg\min_{m\in \mathcal{C}_t}\; E_t(m),
\end{equation}
where $\mathcal{C}_t$ is a candidate pool constructed by indexed matching and staged relaxation, and $E_t$ is an energy that combines musical continuity, harmonic feasibility, style fidelity, and diversity control.

\paragraph{Indexing and Retrieval Signatures.}
We pre-index the dataset by measure to avoid brute-force scanning at inference time.
For each source measure we store a retrieval signature keyed by its measure length (beats) and its discrete attributes, and we cache the associated accompaniment pattern for fast retrieval.
In our final setting we operate in a \emph{reharmonization match mode}: rather than matching exact chord roots, we match the \emph{chord-quality sequence} (e.g., \texttt{maj}, \texttt{min}, \texttt{dom}, \texttt{sus4}) at the beat level.
This choice is musically motivated: chord quality largely determines which scale degrees are stable targets for voicing (triad vs.\ seventh vs.\ dominant vs.\ suspension), while absolute roots can be changed later by reharmonization.
It is also practically important: strict root-level matching collapses candidate pools in POP909, whereas quality-level matching preserves stylistic diversity while keeping harmonic intent legible.
Although the planner is inventory-constrained, we additionally apply a retriever-side \emph{style snapping} safeguard: we project the planned style vector onto the nearest style vector observed in the corpus under a weighted Hamming metric, optionally keeping a subset of slots fixed (\emph{pins}).
This guarantees that indexing is never asked to retrieve an unsupported slot combination (e.g., if a slot is missing or inferred), and it provides a stable ``nearest feasible'' behavior in edge cases.

\paragraph{Chord-Quality Matching as Soft Constraints.}
Even in a reharmonization setting, chord quality still matters for both voice-leading and plausibility.
We represent each measure's harmony by a per-beat quality label sequence $\mathbf{q}=(q_1,\ldots,q_L)$ of length $L$ beats.
Given target labels $\mathbf{q}_t$ and a candidate's labels $\mathbf{q}(m)$, we compute a beatwise mismatch profile and convert it into a penalty:
\begin{equation}
\begin{split}
E_{\text{qual}}(m) \;=\;&
\alpha_{\text{wild}} n_{\text{wild}}
+ \alpha_{\text{miss}} n_{\text{miss}}
+ \alpha_{\text{fam}} n_{\text{fam}} \\
&+ \alpha_{\text{sus}} n_{\text{sus}}
+ \alpha_{\text{cross}} n_{\text{cross}} \\
&+ \alpha_{\text{mismatch}} n_{\text{mismatch}},
\end{split}
\end{equation}
where each count tallies a different relaxation event: (i) \emph{wildcards} (beats for which the quality constraint is ignored), (ii) missing quality labels, (iii) \emph{family matches} (e.g., \texttt{maj}$\leftrightarrow$\texttt{maj7}), (iv) suspension-related matches (extra penalty for dominant-beat suspensions), (v) cross-family matches (used only when necessary), and (vi) true mismatches.
This scoring view lets us widen candidate pools without discarding harmonic intent: exact matches are preferred, but musically adjacent colors are allowed with controlled cost.

\paragraph{Staged Relaxation with Structural Awareness.}
Candidate pool construction is a sequence of increasingly permissive constraints.
We begin with an intersection of (a) planned style $\mathbf{s}_t$, (b) strict quality matches, and (c) structural role constraints (phrase/boundary compatibility).
If the pool is too small, we relax in stages: allow wildcard beats for interior measures (typically on weak beats), allow chord-quality family matches (and then limited cross-family matches for non-boundary interior measures), and finally broaden from \emph{style+quality} matching to \emph{style-only} matching.
When dataset coverage remains insufficient, we add a \emph{bounded style relaxation} stage that includes measures within a small weighted Hamming distance of the planned style vector, preserving the intended ``family'' of accompaniment while increasing feasibility.
For boundary measures (song start/end), we apply stricter role filters and fall back to boundary-aware candidate pools to avoid implausible pickups and cadences.
These stages ensure that retrieval is robust: the system prefers faithful matches but degrades gracefully when the corpus cannot satisfy all constraints simultaneously.
If the fully relaxed pools are empty, we fall back to a neighbor-style strategy (using the previously selected measure or an adjacent measure's style as a proxy) to preserve continuity.
Finally, if a non-empty pool exists but all candidates are rejected during reharmonization (e.g., due to pitch-shift limits), we perform a conservative forced selection among feasible candidates, providing an always-return guarantee.

\paragraph{Reharmonization as Constrained Pitch Mapping.}
Given a selected source measure, we apply reharmonization to align pitches with the target chord progression while preserving rhythmic structure and voice ordering.
We represent target harmony at each beat $b$ as a mask over allowed pitch classes (chord tones), and optionally a key-conditioned diatonic scale set $\mathcal{K}$.
For each onset pitch $p$ in the retrieved pattern that occurs on beat $b$, we choose a mapped pitch
\begin{equation}
p' \;=\; \arg\min_{x \in \mathcal{A}_b}\, |x - p|,
\end{equation}
where $\mathcal{A}_b$ is the beat-specific allowed set.
On beats that require strict alignment (e.g., hard mismatches, wildcard-enforced beats), we set $\mathcal{A}_b$ to target chord tones.
On beats where chord-quality matches permit color tones, we allow a controlled union of chord and candidate color pitch classes.
Otherwise, we treat the retrieved pattern's chord-tone vs.\ non-chord-tone identity as part of its pianistic gesture: chord tones map to target chord tones, while passing tones map to diatonic scale tones.
We reject candidates whose reharmonization would require excessive mean absolute pitch shift or an implausible global register drift, and we enforce a minimum chord-tone ratio to avoid ``scale soup'' under reharmonization.
This mapping is intentionally lightweight, but it preserves a key musical invariant: the \emph{rhythmic motor} and the \emph{voicing topology} come from a real pianist, while harmony is made consistent with the lead sheet.

\paragraph{Continuity: Voice Leading, Anticipation, and Boundary Roles.}
To select among feasible candidates, we optimize continuity across measures.
Let $\mathbf{p}_{t-1}$ and $\mathbf{p}_t(m)$ denote the vectors of first-onset pitches per voice in the previous selected measure and the reharmonized candidate.
We compute a voice-leading cost that approximates pianistic constraints:
\begin{equation}
\begin{split}
E_{\text{vl}}(m) \;=\;&
\sum_v | \Delta p_v |
+ \lambda_{\text{leap}}\!\sum_v \mathbf{1}[|\Delta p_v|>\tau] \\
&+ \lambda_{\text{rest}}\!\sum_v \mathbf{1}[p_v\ \text{missing}] \\
&+ \lambda_{\text{cross}} n_{\text{cross}}
+ \lambda_{\text{par}} n_{\text{par}},
\end{split}
\end{equation}
where $\Delta p_v$ is the voice motion, and we penalize large leaps, voice rests (missing onsets), voice crossings, and parallel fifth/octave motion between adjacent voices.
We also include an \emph{anticipation penalty} that discourages pitch-class or register collisions between sustained notes entering the measure boundary and the candidate's beat-1 onsets.
When the global key is available and the target is the final measure, we also bias selection toward tonic-resolving bass motion to support plausible cadential closure.
Finally, we incorporate role-aware penalties that discourage using ``cadential'' source measures in the middle of a phrase (and vice versa), reflecting that accompaniment patterns are often function-specific in human performances.

\paragraph{Style Fidelity: Distance, Slack Budgets, and Drift Control.}
Beyond matching the discrete planned slots exactly, we treat style as a soft constraint with diagnostic structure.
We compute a style distance either (i) directly on the discrete slot vector (weighted Hamming distance) or (ii) between \emph{style distributions} when the planner supplies a weighted set of candidate style vectors (enabling uncertainty-aware retrieval without text modeling).
We cap the resulting penalty to prevent any single axis from dominating the objective.
To respect the musical semantics of keywords, we also implement targeted guards; for example, when the plan requests \texttt{quiet}+\texttt{gentle}, we reject candidates whose onset activity is above the dataset median, since ``quiet'' accompaniment is not only soft but also typically less busy.

Because no dataset offers perfect coverage of every possible style/harmony/role combination, we allow a controlled number of style deviations via a \emph{slack budget}.
At the phrase and song scales, we permit up to $\lceil \rho_{\text{phrase}} T\rceil$ and $\lceil \rho_{\text{song}} T\rceil$ deviating measures after $T$ selections.
In our current configuration, deviations are only permitted at annotated harmonic transitions (and never at song boundaries), so departures from the planned style are musically motivated rather than arbitrary.
Once this budget is exhausted, deviating candidates are rejected.
In addition, we add a \emph{drift penalty} that monitors running medians of continuous style features (velocity median, staccato ratio, onset rate) and discourages the retrieved accompaniment from drifting outside quantile-derived target bands implied by the planned labels.
This acts as a lightweight ``style thermostat'': it does not enforce exact distributions, but it prevents slow accumulation of unintended intensity or articulation over long form.

\paragraph{Diversity-Aware Selection and Stochastic Decoding.}
Long-form accompaniment fails most often by local repetition: the system finds a ``safe'' pattern and reuses it too frequently.
We therefore add explicit diversity control on the retrieved \emph{pattern signature}, a hash of the active-grid rhythm/texture skeleton.
We penalize consecutive reuse beyond an allowance, apply a cooldown window that discourages recent signatures, and increase penalties when harmonic context changes (so repeated patterns do not flatten harmonic contrast).
Importantly, these penalties are \emph{adaptive}: we scale the repetition penalty based on the observed repeat ratio and unique ratio so far, increasing pressure toward novelty when diversity falls below a target level.
Complementary to signature-level repetition, we also track \emph{measure reuse} and penalize selecting the same source measure repeatedly within the current phrase, section, and song; at phrase/section boundaries we optionally enforce novelty by temporarily blocking measures used in the previous phrase/section, reverting to penalties only when the corpus is too small.

Selection then proceeds in two phases.
First, we gate candidates to preserve musical continuity: among high-quality candidates we keep those whose voice-leading cost lies within a small window of the best candidate, and we deduplicate by pattern signature so that the final pool represents distinct textures.
Second, we choose from the top-$k$ candidates either deterministically (MAP) or stochastically using a Boltzmann distribution,
\begin{equation}
P(m \mid t)\;\propto\;\exp\!\left(-E_t(m)/T\right),
\end{equation}
where temperature $T$ controls exploration.
This yields a simple, musician-tunable diversity knob while maintaining voice-leading and harmonic feasibility.

Overall, the retriever implements a structured optimization that is directly interpretable in musical terms (harmony, role, voice leading, and repetition), while remaining grounded in real pianist performance.
The design is inspired by earlier template-based arrangement systems \cite{zhao2021accomontage} but operates at the measure level with explicit style planning, staged relaxation, and diversity-aware selection tuned for long-form accompaniment.

\section{Experiments and Results}
\label{sec:experiments}

We evaluate the end-to-end pipeline, the planner plus retriever, using \emph{test songs} that are entirely outside POP909 and were never used for training or indexing. Five songs are used for end-to-end diversity and retriever-health analysis. We additionally run a 12-keyword suite on each song to evaluate inter-style isolation and realized style compliance in the generated MIDI.
For evaluation, we provide a demo webpage with 60 MIDI files generated in our experiments, including interactive piano-roll visualizations, at \url{https://piano-paper-demos.github.io/pianocomp-demo/}.

\subsection{Transformer Size and Training}
\label{subsec:planner_training}
The transformer size is designed for a small dataset like POP909. It uses $d_{\text{model}}=256$ with 4 layers and 8 attention heads. The feedforward width is $d_{\text{ff}}=1024$. We train with a masked-slot objective over six categorical, measure-level style slots covering dynamics, articulation, rhythm, tension, texture, and register \cite{devlin2019bert}. Training runs for 20 epochs with learning rate $3{\times}10^{-4}$ and dropout $0.1$. When keyword conditioning is enabled, the prompt-constraint weight is $0.2$. The model converges quickly. The best validation loss is 0.709 at epoch 5 and the best validation slot accuracy is 0.636 at epoch 9. At epoch 20, per-slot validation accuracy is highest for dynamics at 0.783. Accuracy for rhythm, tension, and articulation is 0.651, 0.641, and 0.609. Texture and register are more ambiguous at 0.468 and 0.446. Thus, these results motivate retrieval-side feasibility safeguards. Training and inference use a 32-measure window for style history and a one-measure look-ahead window for music continuity, e.g., to support anticipation.

\paragraph{Interpreting Style-Slot Accuracy.}
The reported validation accuracy should not be interpreted as a ceiling on generation quality. For a given harmonic and structural context, multiple style realizations are often equally valid (e.g., several texture or register choices within a chorus). As a result, the style labels derived from the dataset are inherently multi-modal, and exact classification accuracy underestimates the planner’s utility. In our system, partially correct or uncertain style predictions are stabilized by inventory-constrained decoding and retriever-side feasibility checks, which project planner outputs onto empirically observed style combinations. Consequently, moderate per-slot accuracy is sufficient to yield consistent long-form realizations when coupled with retrieval-based grounding.

\subsection{Pattern Diversity}
We evaluate pattern diversity and retrieval health using four signature-based metrics. These include unique-pattern ratio, defined as unique signatures divided by length, dominant-cluster ratio, repeat ratio for consecutive repeats, and strict-pool miss count when the strict style+QUAL pool is empty and staged relaxation is required.

For context, we compute the same signature-based metrics over POP909 by clustering identical canonical signatures. We find high intrinsic variety with mean unique ratio 0.854 and mean dominant-cluster ratio 0.0329, and rare consecutive repeats with mean repeat ratio 0.000754.

\begin{table}[t]
\centering
\caption{Section-wise diversity and retrieval health. Ent$_\mathrm{norm}$ is normalized entropy. Top-3 is top-3 coverage. Dom. is dominant-cluster ratio. Pool is the median strict candidate pool size for style+QUAL. Miss is the count of measures with an empty strict pool, so relaxation is used.}
\label{tab:five_song_diversity}
{\setlength{\tabcolsep}{2pt}%
\begin{tabular}{@{}llllll@{}}
\toprule
Song & Ent$_\mathrm{norm}$ & Top-3 & Dom. & Miss & Pool \\
\midrule
dngl & 1.000 & 0.399 & 0.166 & 5 & 12 \\
gsn & 1.000 & 0.218 & 0.073 & 6 & 29 \\
lbkch & 0.996 & 0.274 & 0.110 & 4 & 13 \\
rsj & 1.000 & 0.297 & 0.132 & 2 & 31 \\
ts & 0.750 & 0.352 & 0.284 & 3 & 31 \\
\midrule
Mean $\pm$ Std & $0.949\pm0.100$ & $0.308\pm0.063$ & $0.153\pm0.072$ & $4.0\pm1.4$ & $23.2\pm8.8$ \\
\bottomrule
\end{tabular}}
\end{table}

Whole-song ratios are near-saturated. Mean unique ratio is 0.998, mean dominant-cluster ratio is 0.011, and repeat ratio is 0.000. Therefore, we report more sensitive section-wise metrics in Table~\ref{tab:five_song_diversity}. Strict pools are empty in 3.3\% of measures, which is 20/599, and misses trigger staged relaxation without missing output. Repetition can be tuned via $P_{rep}$ in Section~\ref{sec:retriever}. 

Overall, the high section-wise entropy and low dominance indicate that retrieval does not collapse to a small set of templates on test songs. Strict match pools are typically non-trivial, with a median of $23.2$ candidates, supporting diverse, structure-aware realizations in the small-data regime. The results align with real pianist performances and demonstrate high diversity of per-measure patterns while maintaining coherent styles for a whole song. 

\subsection{Ablation Experiments}

We also evaluate planner ablations on the same five test songs, under a fixed retriever. We compare four conditions. A1 uses a flat majority style. A2 uses the style-transformer planner with \texttt{auto}, i.e., the dataset distribution. A3 uses a flat ``first-measure'' baseline that freezes the initial planned style. A1-Rare uses a rare flat style \texttt{quiet+gentle+slow+steady+block+warm}.

In pattern space, diversity metrics are saturated across all conditions, with unique ratio $\approx 0.996$--$0.998$, dominant-cluster ratio $\approx 0.011$, and repeat ratio $\approx 0$. Thus, they do not distinguish planner vs.\ flat baselines. A1 and A1-Rare are also indistinguishable on these aggregates. Both have mean unique ratio 0.99798, dominant-cluster ratio 0.01097, and repeat ratio 0.0000. Their retrieved patterns can still differ substantially.

To better reflect planner intent, we compute style-space metrics from the per-measure target style labels and section annotations. Only the planner condition shows non-zero style motion and boundary-aware contrast. Table~\ref{tab:ablation_style_space} summarizes these effects and reports bootstrap confidence intervals for the planner deltas A2$-$A1 across five songs. Flat baselines are constant in style space. Transition distance is zero and section-mode ratios are one, even when the chosen flat style is rare. Style compliance differs across conditions. Flat styles are easy for the corpus to satisfy, with match rates $\approx 0.99$. The planner targets are more diverse and therefore harder to satisfy, with match rate $\approx 0.81$. The first-measure baseline collapses compliance to near zero. Overall, these ablations suggest that the planner primarily contributes to long-range style motion, while pattern-level diversity ratios alone can be maintained even on near-unique style tokens.

\begin{table*}[t]
\centering
\setlength{\tabcolsep}{2pt}
\caption{Ablation results. We compare a fixed-style baseline without a planner against a learned style planner using style-space metrics derived from target style labels and section structure. Numbers are mean $\pm$ std across songs. We also report 95\% bootstrap CIs with n=5 songs for $\Delta=\text{planner}-\text{fixed}$. All CIs exclude zero, indicating statistically reliable planner effects in style space.}
\label{tab:ablation_style_space}
\begin{tabular}{@{}l p{0.28\linewidth} l l p{0.18\linewidth}@{}}
\toprule
Metric & Definition, brief & Fixed style & Style planner & $\Delta$, 95\% CI \\
\midrule
Transition distance $\uparrow$ & Mean adjacent-measure style change. & $0.000\pm0.000$ & $0.309\pm0.019$ & $+0.309$ $[0.294,0.326]$ \\
Boundary contrast $\uparrow$ & Boundary dist. $-$ within-section dist. & $0.000\pm0.000$ & $0.223\pm0.036$ & $+0.223$ $[0.195,0.259]$ \\
Axis entropy $\uparrow$ & Mean per-axis entropy within sections. & $0.000\pm0.000$ & $0.720\pm0.044$ & $+0.720$ $[0.681,0.761]$ \\
Vector entropy $\uparrow$ & Entropy of full style vectors within sections. & $0.000\pm0.000$ & $3.116\pm0.326$ & $+3.116$ $[2.817,3.395]$ \\
Section mode ratio $\downarrow$ & Share of dominant style vector per section. & $1.000\pm0.000$ & $0.408\pm0.057$ & $-0.592$ $[-0.646,-0.547]$ \\
Axis consistency $\downarrow$ & Mean per-axis mode share within sections. & $1.000\pm0.000$ & $0.758\pm0.024$ & $-0.242$ $[-0.264,-0.221]$ \\
Vector consistency $\downarrow$ & Mode share of full style vectors within sections. & $1.000\pm0.000$ & $0.312\pm0.046$ & $-0.688$ $[-0.732,-0.653]$ \\
\bottomrule
\end{tabular}
\end{table*}

\subsection{Style Isolation and Compliance}
Across the five test songs, keyword prompting yields distinct planned style sequences. Mean pairwise plan differences are $0.940\pm0.006$ of measures. We then evaluate whether retrieval preserves this separation in realized patterns.

We define two per-measure signatures. The first is an \emph{active signature} that captures the rhythm and texture skeleton on the beat grid. The second is a stricter \emph{pitch signature} that additionally encodes pitch content. We report two metrics. The first is an \emph{all-match ratio}, where all 12 runs are identical. The second is a \emph{pairwise match mean}. All-match ratios are 0.0000 for all songs, and Table~\ref{tab:style_isolation_5songs} shows low pairwise overlap with large unions, indicating strong inter-style isolation under keyword control.

\begin{table*}[t]
\centering
\setlength{\tabcolsep}{3pt}
\caption{Inter-style isolation. Pairwise is the mean pairwise match. Union is the number of distinct signatures within the song. The last row reports mean $\pm$ std across songs.}
\label{tab:style_isolation_5songs}
\begin{tabular}{@{}llllll@{}}
\toprule
Song & Measures & Active pairwise & Active union & Pitch pairwise & Pitch union \\
\midrule
dngl & 79 & 0.0215 & 672 & 0.0209 & 770 \\
gsn & 173 & 0.0214 & 969 & 0.0215 & 1191 \\
lbkch & 99 & 0.0223 & 826 & 0.0220 & 963 \\
rsj & 127 & 0.0137 & 960 & 0.0137 & 1156 \\
ts & 121 & 0.0199 & 949 & 0.0203 & 1106 \\
\midrule
Mean $\pm$ Std & $119.8\pm31.6$ & $0.0198\pm0.0031$ & $875.2\pm114.1$ & $0.0197\pm0.0030$ & $1037.2\pm154.5$ \\
\bottomrule
\end{tabular}
\end{table*}

Table~\ref{tab:style_isolation_5songs} shows consistently low overlap across all songs, supporting strong inter-style isolation under keyword control. Different keywords yield distinct realized patterns at most measure positions, even under a shared structural scaffold.

To verify that keyword control affects the realized accompaniment, rather than only discrete slot labels, we evaluate continuous performance features from the rendered MIDI across the five test songs. We use time-aligned measure boundaries. Table~\ref{tab:kw_realized_5songs} reports note-start rate, median velocity, staccato ratio for dur$\leq 0.25$ beat, and register span computed as P95$-$P5. Numbers are mean $\pm$ std across songs.

\begin{table}[t]
\centering
\caption{Realized continuous features.}
\label{tab:kw_realized_5songs}
{\setlength{\tabcolsep}{2pt}%
\begin{tabular}{@{}lllll@{}}
\toprule
Style & Starts/Meas & Vel$_\mathrm{med}$ & Staccato & RegSpan \\
\midrule
arp & $9.51\pm0.33$ & $73.0\pm4.4$ & $0.093\pm0.026$ & $25.0\pm1.3$ \\
block & $9.19\pm0.39$ & $73.4\pm5.2$ & $0.111\pm0.045$ & $26.6\pm1.7$ \\
busy & $15.36\pm0.76$ & $75.4\pm4.3$ & $0.287\pm0.054$ & $26.2\pm1.6$ \\
dense & $15.19\pm1.18$ & $74.0\pm4.7$ & $0.228\pm0.055$ & $25.4\pm1.5$ \\
fast & $16.74\pm0.86$ & $73.8\pm4.4$ & $0.230\pm0.050$ & $26.2\pm1.6$ \\
gentle & $11.18\pm0.86$ & $74.4\pm5.1$ & $0.057\pm0.014$ & $25.6\pm1.5$ \\
loud & $13.96\pm1.15$ & $92.2\pm0.4$ & $0.226\pm0.051$ & $26.4\pm2.0$ \\
ostinato & $9.39\pm0.46$ & $74.6\pm4.4$ & $0.104\pm0.025$ & $24.4\pm0.8$ \\
quiet & $11.13\pm0.83$ & $66.2\pm0.7$ & $0.054\pm0.005$ & $27.8\pm0.7$ \\
slow & $8.50\pm0.52$ & $74.2\pm5.6$ & $0.055\pm0.013$ & $25.2\pm1.9$ \\
sparse & $8.30\pm0.44$ & $73.8\pm5.2$ & $0.045\pm0.011$ & $26.8\pm1.7$ \\
stride & $9.00\pm0.48$ & $72.8\pm3.2$ & $0.060\pm0.025$ & $27.8\pm1.0$ \\
\bottomrule
\end{tabular}}
\end{table}

Table~\ref{tab:kw_realized_5songs} shows that prompts shift realized features in the expected directions, including activity, dynamics, articulation, and register. Together with the low overlap in Table~\ref{tab:style_isolation_5songs}, this indicates that keyword control changes both discrete realizations and measurable performance-level properties after retrieval and reharmonization.
\section{Discussion}
\label{sec:discussion}

Retrieval-based generation provides highly realistic note-level realizations by reusing patterns played by real pianists, including articulation and voicing idioms. However, diversity is ultimately bounded by the corpus. POP909 contains 77,638 measures, which appears large at first glance. However, the effective pool is constrained by multiple factors. Accompaniment style is relatively consistent within each song. POP909 contains only 909 songs total. Measures occupy specific structural roles. We do not mix certain musical categories such as mode. POP909 has 503 major and 406 minor songs. We also do not mix meter. POP909 has 880 songs in 4/4 and only 29 in 3/4 or 6/8. Moreover, because POP909 is curated from Chinese pop songs, the corpus exhibits stylistic bias.

Taken together, these constraints make 77,638 measures less abundant than the raw count suggests. This is reflected by the strict candidate-pool sizes and miss counts in Table~\ref{tab:five_song_diversity}. In our test set, strict matching can fail in 3.3\% of measures, meaning the strict style+QUAL pool is empty. This requires staged relaxation to retrieve a fallback pattern. Currently, we use close styles such as \texttt{quiet} for \texttt{gentle} and chord quality family matching such as \chordtype{Dm7} for \chordtype{Dm} for relaxed matching. We do this to maintain musical quality of retrieval choices.

\subsection{Limitations}

Therefore, the main limitation of our approach is corpus-limited diversity, including finite size and domain-specific stylistic coverage. In addition, our outputs are symbolic MIDI. High-quality audio rendering and perceptual evaluation in end-to-end settings are beyond the scope of this work.

Our approach has several limitations.
First, retrieval bounds creativity: the system cannot invent patterns outside the dataset, and rare styles are constrained by small candidate pools.
Second, reharmonization preserves rhythm and voicing but may introduce artifacts in edge cases (e.g., altered chords or non-chord tones) because it must map pitches into a new harmonic context.
Third, our current style planner predicts measure-level style slots; beat-level dynamics and within-measure texture variation are not modeled explicitly.
Finally, our evaluation focuses on pattern diversity and retrieval health; complementary perceptual studies are needed to assess musical quality in listening tests.

Overall, we view this work not as a replacement for end-to-end music generation models, but as evidence that explicit planning and dataset-aligned retrieval can serve as a strong and interpretable alternative when controllability, structure, and small-data robustness are primary concerns.

\section{Future Work}
\label{sec:futurework}

A natural extension is to predict finer-grained controls, including beat-level dynamics and intra-measure texture changes, and to couple these controls to reharmonization and selection.
Hybrid systems are another direction: a lightweight neural decoder could edit or inpaint retrieved measures to better fit difficult harmonic contexts while retaining retrieval's grounding.
Expanding the retrieval dataset beyond POP909 and studying cross-dataset transfer would further test robustness and improve coverage for rare textures.

\section{Conclusion}
\label{sec:conclusion}

We presented a structure-aware piano accompaniment system that decouples style planning from note-level realization. A compact Transformer predicts an interpretable sequence of measure-level style slots conditioned on functional harmony and section/phrase structure, and a dataset-aligned retriever selects and reharmonizes real pianist patterns with voice-leading and diversity-aware constraints.

Our experiments on songs outside POP909 show strong inter-style isolation under keyword prompting and robust long-form behavior under retrieval constraints, while planner ablations confirm the role of learned style planning in producing coherent style motion. Overall, retrieval-first accompaniment provides a practical path to controllable long-form output in the small-data regime.
\section*{Acknowledgments}

We thank the creators of the POP909 dataset for making the performances and annotations publicly available. We used the Codex CLI coding assistant from OpenAI~\cite{openai} to accelerate implementation.

\bibliographystyle{IEEEtran}
\bibliography{refs}

\section*{Appendix}

\subsection*{Transformer Planner Training}

\paragraph{Training Configuration and Results}
Key parameters for the transformer style planner include: learning rate 3e-4, past window size 32, future window size 1, and drop out rate 0.1. When prompt conditioning is enabled, we apply soft prompt constraints with prompt constraint weight 0.2.

\begin{figure}[t]
\centering
\begin{tikzpicture}
\begin{groupplot}[
    group style={group size=1 by 2, vertical sep=1.1em},
    width=\columnwidth,
    height=0.42\columnwidth,
    xmin=1, xmax=20,
    xtick={1,5,10,15,20},
    grid=both,
    grid style={line width=.1pt, draw=gray!30},
    major grid style={line width=.2pt, draw=gray!45},
    legend style={draw=none, fill=none, font=\scriptsize, cells={anchor=west}},
    legend pos=north east,
]
\nextgroupplot[
    ylabel={Loss},
    ymin=0.55, ymax=0.82,
]
\addplot+[mark=*, mark size=1pt] coordinates {
    (1,0.7730) (2,0.7250) (3,0.7140) (4,0.7081) (5,0.7040)
    (6,0.6998) (7,0.6963) (8,0.6919) (9,0.6877) (10,0.6828)
    (11,0.6772) (12,0.6707) (13,0.6633) (14,0.6545) (15,0.6449)
    (16,0.6339) (17,0.6220) (18,0.6084) (19,0.5942) (20,0.5787)
};
\addlegendentry{train}
\addplot+[mark=square*, mark size=1pt] coordinates {
    (1,0.7304) (2,0.7161) (3,0.7165) (4,0.7102) (5,0.7090)
    (6,0.7109) (7,0.7153) (8,0.7157) (9,0.7116) (10,0.7135)
    (11,0.7209) (12,0.7196) (13,0.7233) (14,0.7353) (15,0.7399)
    (16,0.7492) (17,0.7591) (18,0.7602) (19,0.7886) (20,0.7988)
};
\addlegendentry{val}

\nextgroupplot[
    ylabel={Accuracy},
    xlabel={Epoch},
    ymin=0.58, ymax=0.73,
    legend pos=south east,
]
\addplot+[mark=*, mark size=1pt] coordinates {
    (1,0.5887) (2,0.6202) (3,0.6261) (4,0.6304) (5,0.6325)
    (6,0.6354) (7,0.6378) (8,0.6407) (9,0.6430) (10,0.6465)
    (11,0.6504) (12,0.6545) (13,0.6598) (14,0.6648) (15,0.6711)
    (16,0.6782) (17,0.6861) (18,0.6939) (19,0.7027) (20,0.7120)
};
\addlegendentry{train}
\addplot+[mark=square*, mark size=1pt] coordinates {
    (1,0.6301) (2,0.6301) (3,0.6317) (4,0.6334) (5,0.6352)
    (6,0.6348) (7,0.6246) (8,0.6306) (9,0.6363) (10,0.6353)
    (11,0.6308) (12,0.6329) (13,0.6283) (14,0.6263) (15,0.6237)
    (16,0.6189) (17,0.6161) (18,0.6101) (19,0.6041) (20,0.5995)
};
\addlegendentry{val}
\end{groupplot}
\end{tikzpicture}
\caption{Training dynamics of the Transformer style planner (20 epochs). The model converges quickly, with validation loss and accuracy peaking early.}
\label{fig:planner_curves}
\end{figure}
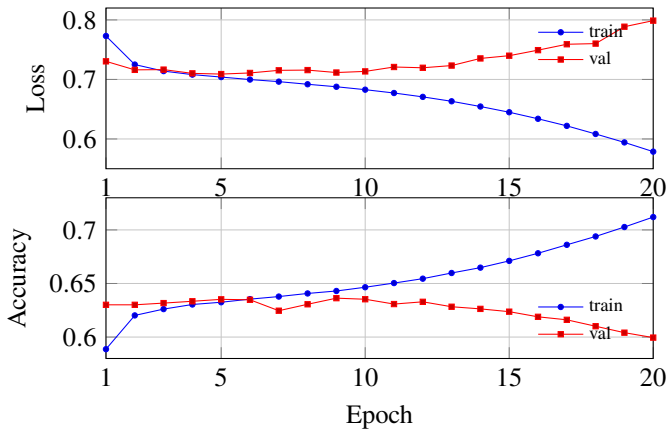

Fig.~\ref{fig:planner_curves} plots training and validation metrics for the run reported in Section~\ref{subsec:planner_training}. Validation metrics peak early, supporting checkpoint selection and early stopping in the small-data regime.

\subsection*{Ethical Consideration}

This work investigates assistance for symbolic arrangement rather than automated replacement of human composers or performers.
Our system is designed to keep users in control through transparent intermediate representations (style slots) and dataset-grounded retrieval, so that musical decisions remain interpretable and editable.
We do not collect personal data, and we do not conduct human-subject experiments in this paper.
As with any data-driven creative tool, we encourage appropriate attribution and responsible use of generated outputs in downstream creative or commercial settings.

\subsection*{Dataset Licensing}

We use the POP909 dataset and comply with the license terms provided by its official release repository (MIT license).
Our experiments operate on locally processed derivatives (e.g., indices and feature representations) for research evaluation; we do not redistribute POP909 files as part of this paper.
Any release of code and processed artifacts should preserve attribution and remain consistent with the dataset's licensing terms.

\end{document}